\documentclass[12pt]{iopart}
\usepackage{iopams} 
\usepackage[dvips]{epsfig}

\def\as{{\rm{asymp}}}

\def\bra#1{ \langle{#1}| }
\def\braket#1#2{ \langle{#1}|{#2}\rangle }
\def\opket#1#2 {  {#1} |{#2}\rangle }
\def\braopket#1#2#3{ \langle{#1}| {#2} |{#3}\rangle }

\def\d{{\rm d}}

\def\dydxh#1#2{ \partial{#1} / \partial {#2} }
\def\dydxv#1#2{ {{\partial #1} \over {\partial #2}} }

\def\dydxvo#1#2{ {{\d #1} \over {\d #2}} }

\def\ds{\displaystyle}

\def\F{{\cal F}}

\def\ha{{1 \over 2}}
\def\H{{\cal H}}
\def\hh{\hat{h}}
\def\Hh{\hat{H}}
\def\I{{\cal I}}
\def\Ih{\hat{I}}

\def\in{{\rm in}}
\def\J{{\bf J}}

\def\K2{{\cal K}}
\def\ket#1{ |{#1}\rangle }

\def\mat#1#2#3#4{  \left( \matrix{ {#1} & {#2} \cr
                                   \noalign{\vskip3pt}
                                   {#3} & {#4} \cr    } \right) }

\def\Nchan{M}

\def\out{{\rm out}}
\def\p{{p}}
\def\phat{\hat{p}}

\def\ph#1{\phantom{#1}}

\def\Ph{\hat{P}}
\def\q{{q}}
\def\qhat{\hat{q}}

\def\Qh{\hat{Q}}
\def\R{\hat{\cal R}}
\def\RC{{\rm rc}}

\def\T{{\hat{\cal T}}}

\def\tr{{\rm tr}}
\def\Tr{{\rm Tr}}

\def\TS{{\rm ts}}

\def\Wt{{\cal W}}

\def\spinor#1#2{\left( \begin{array}{c} {#1}\\[9pt]
			                {#2} \end{array} \right)}

\def\sbar{\big/}
\def\dbar{ {\,\sbar\!\sbar} }
\def\dbarr{ {\,\sbar} }
\def\brasket#1#2#3{ \langle{#1}\dbar_{#2}{#3}\rangle }
\def\opsket#1#2#3 {  {#1} \dbar_{#2}{#3}\rangle }

\def\brassket#1#2#3{ \langle{#1}\dbarr_{#2}{#3}\rangle }
\def\opssket#1#2#3 {  {#1} \dbarr_{#2}{#3}\rangle }

\begin{document}

\title{Semiclassical transmission across 
transition states}

\author{Stephen C. Creagh}
\address {School of Mathematical Sciences, University of Nottingham, 
	Nottingham NG7 2RD, UK.}

\begin{abstract}
It is shown that the probability of quantum-mechanical transmission
across a phase space bottleneck can be compactly approximated using 
an operator derived from a complex Poincar\'e return map. This 
result uniformly incorporates tunnelling effects with 
classically-allowed transmission and 
generalises a result previously derived for a classically small
region of phase space.
\end{abstract}


\section{Introduction} 
\label{introduction}
There has recently been a resurgence of interest in the classical
transition state theory of molecular reactions.
Results that were historically restricted to two degrees of freedom 
\cite{Pechukas,PODS2} have been generalised
to arbitrary dimensions using the construction of normally hyperbolic
invariant manifolds (or NHIM's) \cite{Jaffe}-\cite{LW1}. It is natural 
to ask how classical structure such as the NHIM
is reflected in the quantum-mechanical problem, which corresponds to 
scattering from a multidimensional potential barrier.

An answer to this question has been offered in \cite{me},
where a description is given of quantum mechanical transport across 
a phase-space bottleneck using dynamics linearised around a 
certain complex periodic orbit. In using linearised dynamics
these results are restricted to a classically small region of 
phase space and energies that are no larger than $O(\hbar)$
above a transmission threshold. In this paper it 
is shown how fully nonlinear dynamics may be incorporated in this approach, 
resulting in a description of transport which is not restricted to a 
small region of phase space or energy range above threshold. The current 
approach is based on a quantisation of a classical normal form Hamiltonian, 
although the final form can be expressed in such a way that explicit 
calculation of a normal form is not necessary.

\begin{figure}[h]
\vspace*{0.3in}\hspace*{1.5cm}
\psfig{figure=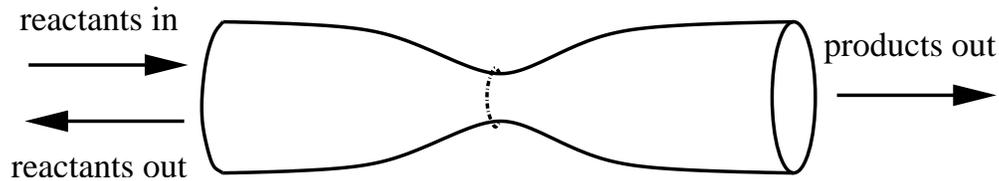,height=0.9in}
\vspace*{0.3in}
\caption{A schematic illustration is given of the waveguide problem considered 
in this paper, showing the case for an energy above threshold in which
the transition state (part of which is indicated schematically by a dashed 
curve) separates reactants from products where the waveguide is at its 
narrowest.
The surface shown is meant to illustrate a level surface in configuration space
of the potential energy in the case of a three-degree-of-freedom Hamiltonian 
that is of the form kinetic-plus-potential (although no such explicit 
assumption regarding the form of the Hamiltonian is made in the calculation).
Note that we use the term ``waveguide'' in a generalised sense in this 
paper to mean a Hamiltonian with a phase space bottleneck which asymptotes
to a vibrational problem decoupled from free motion, and we do not deal 
explicitly with hard-wall or discontinuous potentials.
}
\label{wgfig}
\end{figure}

To describe the result more concretely, let us consider a waveguide problem
with configuration space coordinates $(x,y)$ in which $x$ 
represents longitudinal position along the waveguide and $y$
represents transverse vibrations. If necessary we can let 
$y=(y_1,\cdots,y_d)$ be multidimensional. In chemical applications,
$x$ might be a reaction coordinate with $x$ large and negative
corresponding to decoupled reactant molecules and  $x$ large and positive
corresponding to decoupled product molecules (see Figure~\ref{wgfig}),
while $y$ describes internal vibrations of the reacting molecules.
The quantum mechanics of this problem are described using a 
scattering matrix, which we write in the form
\[
S(E) = \mat{r_{RR}}{t_{RP}}{t_{PR}}{r_{PP}},
\]
where, for example, the block $t_{PR}$ maps asymptotic incoming states
on the reacting side to  the corresponding asymptotic outgoing states
on the product side. In chemical jargon,  $t_{PR}$ gives state-selected
reaction rates (labelled by the incoming mode number) together
with the distribution of product states (labelled by the outgoing
mode number). 
In this paper we will describe a semiclassical approximation
for the operator
\[
\R(E) = t_{PR}^\dagger t_{PR}
\]
which gives a probability of transmission for states incoming on 
the reactant side but which sums over outgoing states and does
not give the distribution of product states. Although containing
less information than the scattering matrix, this {\it reaction operator} 
has clear experimental relevance and, importantly in this context, admits 
semiclassical approximations which are considerably simpler.
As described in detail in \cite{me}, this is because the orbits
used in semiclassical approximation of $S(E)$ are singular near the 
boundary of the reacting subset of phase space whereas those used 
for $\R(E)$ are not.

The reaction operator $\R(E)$ has a clear relationship with the
geometry of the classical transition state. It 
acts on the Hilbert space $\H_R^\in$ of asymptotically propagating
incoming states and the classical analogue of this space
is a Poincar\'e section $\Sigma_R^\in$ obtained by fixing the 
reaction coordinate and the total energy $E$, for which $(y,p_y)$ 
provide canonical coordinates.
Let $E$ be greater than threshold so that there is a nonempty
reacting subset $V$ of $\Sigma_R^\in$ --- the boundary of $V$ is
the intersection with $\Sigma_R^\in$ of the stable manifold of the
NHIM. Then a phase 
space representation of $\R(E)$ such as the Weyl symbol 
$\Wt_{\R}(y,p_y)$ tends to the characteristic function
of the reacting region $V$
\[
\Wt_{\R}(y,p_y) \sim \chi_V(y,p_y)
\]
in the classical limit. Moreover, for finite values of $\hbar$,
$\Wt_{\R}(y,p_y)$ also
incorporates quantum effects such as tunnelling, especially 
important outside $V$ and near its boundary.

An explicit semiclassical approximation was presented for $\R(E)$ in 
\cite{me}, of the form
\begin{equation}\label{uniR}
\R(E) = \frac{\T(E)}{1+\T(E)},
\end{equation}
where the operator $\T(E)$ is constructed from the linear stability properties
of a complex periodic orbit. This complex periodic orbit has a real
initial condition in the interior of $V$ and returns to
it after encircling the transition state region in a 
net imaginary time. The formula was derived by using a separable
approximation of the Hamiltonian in the transition-state region
to match waves propagating in the reactant and product channels.
This separable approximation is valid only insofar as the 
transition state is small on classical scales and the result should
therefore work only when the energy $E$ is within $O(\hbar)$ of 
threshold, where the Liouville volume of $V$ is $O(\hbar^d)$.

We will now show that, as long as the operator $\T(E)$ is interpreted
using fully nonlinear dynamics in a neighbourhood of the periodic orbit, 
Equation~(\ref{uniR}) is in fact valid over classical scales.
The difference between the approach in \cite{me} and the philosophy
applied here has an analogy in the classical treatment of one-dimensional
WKB solutions near turning points. The simplest way to treat turning points
is to approximate the potential using a truncated Taylor expansion
(linear for a single turning point and quadratic for two 
coalescing turning points) and to use the resulting solutions 
to match standard WKB approximations on either side.
The method of comparison equations \cite{BM}, on the other 
hand, seeks a change of variable which (up to higher-order corrections 
in $\hbar$)
maps the potential more globally into a linear or quadratic form as required
and this has the advantage of giving uniform results which apply
over classical length scales. The approach in \cite{me} is analogous
to the method of truncating Taylor series whereas in this publication
we pursue a transformation into normal form that is similar in spirit to
the method of comparison equations.

The difference is that, for multidimensional problems,
a deformation of configuration space variables alone as used in
the method of comparison equations does not have sufficient
range to put the problem in a solvable form and we must use 
transformations in phase space \cite{MillerSmatrix,Cargo,CdVP1}. 
In this way a Hamiltonian which is a quantum analogue of the normal forms in 
\cite{Uzer} can be used and we arrive at a problem which, while not 
separable, is simple enough that scattering solutions can be written down.
More importantly, the information we need to approximate $\R(E)$
can be formulated in such a way that explicit reference to the normal form
can be removed and the end result is a formula (of the same form as 
Equation~(\ref{uniR})) which can be understood simply in terms of complex 
orbits starting and finishing on $\Sigma_R^\in$.

We conclude this section with a brief overview of the paper. The 
essential features of the classical normal form are described in 
Section~\ref{nfsec} and an overview is given of the corresponding 
quantum Hamiltonian. The normal form Hamiltonian is not separable
but does have scattering solutions that are of separable form and these are 
described in Section~\ref{qnfsol}. Because these scattering solutions
do not fully separate in the eigenvalue equation and because the 
normal form Hamiltonian is not of kinetic-plus-potential type, using 
them to describe the transmission properties of a general scattering 
state is not straightforward. Nevertheless a simple solution to this 
problem is possible, which
is outlined in Section~\ref{Fluxsec}. The final step in obtaining a 
usable result is to present the scattering solution in a basis-independent
way, which we do in Section~\ref{nonnfsec} in terms of quantised 
complex Poincar\'e mappings. Conclusions are presented in 
Section~\ref{conclusion}.

\section{Classical and quantum normal forms}\label{nfsec}
The basis for the calculation in this paper is a quantisation 
of the classical normal form for the Hamiltonian around an 
equilibrium. The classical normal form and its connection with 
classical transition state theory are described in detail
in \cite{Uzer}. In this section we will describe the essential results
and adapt some of the notation for our own purposes. We will
then describe the important properties of a quantisation of 
this normal form.

\subsection{The classical normal form}\label{cnfsec}

\begin{figure}[h]
\vspace*{0.3cm}\hspace*{3.5cm}
\psfig{figure=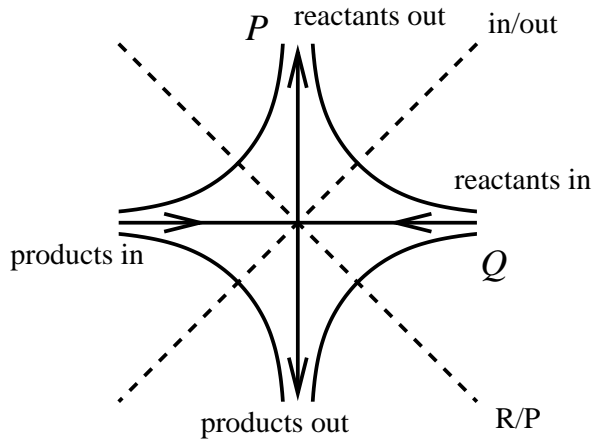,height=2.5in}
\vspace*{0.2in}
\caption{In classical phase space, we can identify in and out 
reactant and product channels with sectors in the $QP$-plane
as illustrated in this figure. Orientations here are for
$\dydxh{H}{I}<0$. The dashed line labelled R/P divides reactants 
from products and the dashed line labelled in/out divides
incoming from outgoing trajectories. For example, a point in phase 
space which projects to the sector in the $QP$-plane between dashed 
lines and containing the positive $Q$-axis is in the incoming 
reactant channel. The hyperbolae illustrate (projections of) typical 
trajectories evolving from the incoming to the outgoing channels.
Note that in the classical picture dynamics is limited to an 
energetically allowed region bounded by hyperbolae in the second and 
fourth quadrants (assuming an energy above threshold), which is
not illustrated here.}
\label{QPfig}
\end{figure}

Let canonical coordinates 
$(q_0,p_0,\q,\p)=(q_0,p_0,q_1,\cdots,q_d,p_1,\cdots,p_d)$, be chosen so 
that the quadratic part of the Hamiltonian is
\[
H(q_0,p_0,\q,\p) = \frac{\lambda}{2}(p_0^2-q_0^2) 
+ \sum_{i=1}^d\omega_i(q_i^2+p_i^2) + {\rm h.o.t.} 
\]
We denote by $f=1+d$ the number of degrees of freedom and we will refer
to $(q_0,p_0)$ as the reaction coordinates. In the context of 
collinear molecular collisions it is useful
to let $q_0=0$ define a dividing surface between reactants and products
with $q_0<0$ corresponding to reactants and $q_0>0$ corresponding to 
products. The reaction coordinates  $(q_0,p_0)$ are useful in 
interpreting the dynamics in this system, in which reaction amounts
to crossing a parabolic potential barrier in that degree of freedom.
It will also be useful for computational purposes however to allow
alternative coordinates $(Q,P)$ defined as a rotation of $(q_0,p_0)$ 
so that
\[
I = \frac{1}{2}(q_0^2-p_0^2)= QP.
\]
We will refer to $I$ as the reaction action. It is positive for 
nonreacting trajectories, which are repelled by the parabolic barrier,
and it is negative for the reacting trajectories, which cross over
(see Figure~\ref{QPfig}).
 
At higher order these canonical coordinates are defined so that
the Hamiltonian depends on the reaction coordinates through the
reaction action $I$ only. To establish notation, we will write
\begin{equation}\label{cnf}
H(Q,P,\q,\p) = H_\RC(I) + H_\TS(\q,\p,I)
\end{equation}
where
\[
 H_\TS(\q,\p,I) =  H_0(\q,\p) + IH_1(\q,\p) +  I^2H_2(\q,\p)+\cdots
\]
The leading parts of  $H_\RC(I)$ and  $H_0(\q,\p)$ contain the quadratic
truncation of the Hamiltonian shown above. That is
\[
 H_\RC(I) = -\lambda I + {\rm h.o.t.}
\]
and
\[
 H_0(\q,\p) =  \sum_{i=1}^d\omega_i(q_i^2+p_i^2) + {\rm h.o.t.}
\]
We refer to  $H_\RC(I)$ and  $H_\TS(\q,\p,I)$ respectively as 
the reaction coordinate part and the transition-state part of 
the Hamiltonian, whence the subscripts.

We remark that for the following calculation to work, it is 
not necessary to put the transition-state part of the Hamiltonian
in normal form. That is, the functions $H_m(\q,\p)$ need not be written
as functions of the transverse actions $J_i=(q_i^2+p_i^2)/2$. This 
allows greater 
latitude in the normal form construction and might in principle alleviate
problems with small denominators. We also remark that since we will
be looking later at complex solutions to the equations of motion,
there is an implicit assumption throughout this paper that the Hamiltonian
is analytic in the transition state region.

\subsection{The quantum normal form}
The transformation to the classical normal form described above is 
achieved by making a canonical change of coordinates so that the
Hamiltonian takes the desired form. The corresponding procedure
in quantum mechanics is to use a unitary change of basis to the 
same effect. While a canonical transformation does not
uniquely define a unitary operator in the quantum formalism, it is 
a well-established feature of the quantum-classical correspondence
that such a connection can be achieved within semiclassical 
approximation. The connection was outlined by Miller in 
\cite{MillerSmatrix} using generating functions to write explicit 
approximations for corresponding unitary operators up to corrections 
of relative order $O(\hbar)$. Explicit and concrete rules describing
how unitary transformations may be constructed which achieve normal 
form to higher order in $\hbar$ have recently been published 
by Cargo et al in \cite{Cargo} and used there to derive compact higher-order
Bohr-Sommerfeld rules. In the context of critical transmission, which
is the application of interest here, transformation to quantum normal 
forms have been exploited by Colin de Verdi\`ere and Parisse in 
\cite{CdVP1,CdVP2,CdVP3} 
to provide connection formulas describing the behaviour of 
wavefunctions near hyperbolic fixed points, and these have been used
in \cite{EBK} to calculate multidimensional quantisation rules which 
are valid near degenerate tori.

An alternative approach is to work directly
with the quantum problem in making the transformation to normal form
rather than simply quantising the classical 
normal form as we do here (see \cite{Ali,Eckhardt,Crehan}, for example). We 
do not adopt this viewpoint because we will in any case later need  
to employ semiclassically constructed nonperturbative unitary 
transformations to connect the normal form basis to the asymptotic 
basis used for the scattering matrix and there is no overall advantage 
in avoiding their use at this stage.

In this work we are interested only in constructing the quantum normal 
form to leading order semiclassically --- that is, neglecting corrections 
of relative order $O(\hbar)$ in wavefunctions or terms of order 
$O(\hbar^2)$ in classical symbols. This is achieved using the 
``preliminary transformation'' in the  language of \cite{Cargo} and the 
resulting leading-order quantum normal form can be written 
straightforwardly as a direct copy of the classical normal form. A 
detailed discussion of this point would unnecessarily 
elongate the presentation here and will simply assert a direct 
equivalence (modulo higher order corrections) between quantum and 
classical Hamiltonians whenever necessary, referring the reader to 
\cite{Cargo,CdVP1} for a proper explanation.

We therefore start, in analogy with (\ref{cnf}), with a Hamiltonian
of the form
\begin{equation}\label{qnf}
\Hh = H_{\RC}(\Ih) + H_\TS(\qhat,\phat,\Ih)
\end{equation}
where $\Ih$ and $H_\TS(\qhat,\phat,\Ih)$ respectively denote 
quantisations of the classical symbols $I$ and $H_\TS(\q,\p,\Ih)$.
There are ordering issues in this correspondence, of course, but
for the purposes of making semiclassical approximation to
leading order in $\hbar$, it suffices to let $\Ih$ and 
$H_\TS(\qhat,\phat,\Ih)$ be Weyl quantisations.
The key feature here is that $\Ih$ and $(\qhat,\phat)$ act 
on different degrees of freedom and therefore commute.
For concreteness, it may occasionally help to suppose that the 
transition-state part can be expanded in the form
\begin{equation}\label{expandH}
H_\TS(\qhat,\phat,\Ih) = H_0(\qhat,\phat) 
+ \Ih H_1(\qhat,\phat) + \Ih^2 H_2(\qhat,\phat) +\cdots
\end{equation}
where we may in particular assume that
\[
[\Ih,H_m(\qhat,\phat)]=0.
\]
The central result in this paper will be stated in an invariant way
that does not refer explicitly to the normal form construction and 
the details of how this transformation is performed are not needed 
to use it. In addition, the essential idea of the calculation is 
understood simply on the basis of the normal form Hamiltonian itself.
We will therefore simply quote the result and refer to 
Refs.~\cite{MillerSmatrix,Cargo,CdVP1} for detailed discussions of 
various approaches to making this transformation in practice.

\section{Scattering solutions of the quantum normal form}\label{qnfsol}
Because higher-order terms in (\ref{expandH}) couple the reaction 
degree of freedom to the transverse degrees of freedom, the normal 
form Hamiltonian is not separable  in the simple-minded sense
of the eigenvalue equation separating into a function of the
reaction coordinate plus a function of the transition-state coordinates.
Since the Hamiltonian depends
on the reaction coordinate only through the reaction action $I$, however, 
it turns out that the eigenvalue equation nevertheless admits
solutions which have a separable structure. We will use this property to 
reduce the 
transmission problem to one that is effectively one-dimensional and 
therefore solvable by standard techniques. Technical details of this 
reduction are given in the present section. In the next, it is shown how the 
results can be formulated in such a way that they no longer rely on 
an explicit consideration of the normal form.

\subsection{The reaction coordinate part}\label{rcsol}

The normal form construction provides us with a 
transformation to coordinates $(Q,P)$ such that $\Ih$ takes 
the form 
\begin{equation}\label{IQP}
\Ih =  \ha\left(\Qh\Ph+\Ph\Qh\right),
\end{equation}
which is the Weyl quantisation of $QP$.
For interpretation of the results below in terms of conventional
calculations, it may help to suppose that 
the reaction coordinate part $\Hh_{\RC}$ of the total Hamiltonian 
acts on functions of a coordinate $x$ (with 
conjugate momentum $p_x$) so that
\begin{equation}\label{Ixp}
\Ih = I(\hat{x},\phat_x)
\end{equation}
and that the problem in the $x$-representation is close to a standard 
barrier-penetration problem. We could, for example, let $(x,p_x)$
coincide with the coordinates $(q_0,p_0)$ defined in Section~\ref{cnfsec}
as a rotation of $(Q,P)$ in the reaction-coordinate phase plane.
In that case the transformation
from (\ref{Ixp}) to (\ref{IQP}) is achieved using a metaplectic 
rotation which rotates the phase plane clockwise through 
an angle $3\pi/4$ (to give  Figure~\ref{QPfig}).

The operator $\Ih$ has continuous spectrum and in $x$-representation
we write the (improper) eigensolutions in the form
\[
\Ih\psi_\I(x) = \I\psi_\I(x).
\]
Then
\[
E_\RC(\I) = H_\RC(I=\I)
\]
is the corresponding reaction-coordinate energy. These 
eigenfunctions are two-fold degenerate since we can send 
incoming waves from either the reactant or the product side
of the barrier (see \ref{1d}). We will restrict our attention here to states 
which have an incoming component on the reactant side and outgoing 
components on both the reactant and product sides, but no incoming 
component on the product side, in which case there is a unique 
solution for each $\I$.

Either as an inverted parabolic barrier \cite{BM} or in the representation
implied by (\ref{IQP}) \cite{CdVP1}, the problem of finding eigensolutions of 
$\Ih$ can be solved exactly and solutions of the scattering problem 
written in closed form. Details
are given in \ref{1d}. For present purposes it is 
sufficient to note that there is a simple relationship describing
the relative fluxes in the incoming and outgoing channels. Let the
scattering state $\psi_\I(x)$ be normalised so that the incoming
flux on the reactant side is normalised to unity (by construction,
the incoming flux on the product side is zero). Then the 
outgoing fluxes on the reactant and product sides are, respectively,
\begin{equation}\label{TR}
T(\I) 
=\frac{1}{1+\e^{2\pi\I/\hbar}}
\qquad\mbox{and}\qquad R(\I) = \frac{1}{1+\e^{-2\pi\I/\hbar}}.
\end{equation}
In the barrier-penetration picture, $T(\I)$ and $R(\I)$ respectively
represent probabilities of transmission and reflection. Note that by 
writing the transmission probability in the alternative form
\[
T(\I) = \frac{\e^{-2\pi\I/\hbar}}{1+\e^{-2\pi\I/\hbar}}
\]
the unitarity condition
\[
 R(\I) + T(\I) =1
\]
becomes self-evident.

We make the following observations concerning this result.
\begin{itemize}
\item There is a symmetry between $R(\I)$ and $T(\I)$ on changing
the sign of $\I$. This is to be expected on the basis of a 
phase-space portrait in $(Q,P)$ coordinates (see Figure~\ref{QPfig}) 
in which changing the sign of $\I$ simply exchanges reactants for products 
in the outgoing channels, but is less obvious in a barrier-penetration 
picture.
\item The transmission and reflection coefficients do not change
if we replace $\Ih$ by a Hamiltonian $H(\Ih)$ which is an arbitrary
function of $\Ih$. This will be obvious after generalised fluxes are 
defined in the next section and a multidimensional version of this 
observation will be important in getting a simple formulation of the 
results in this paper.
\item The expressions in (\ref{TR}) give semiclassical approximations
to transmission and reflection coefficients for a generic potential 
barrier and can be derived from the standard representation of the 
Schr\"odinger equation using the method of comparison equations \cite{BM}.
In the current calculation they are exact for any Hamiltonian which 
can be written as a function of the reaction action $\Ih$ alone. However,
there is in general semiclassical error arising from the transformation 
to normal form in the first place, during which terms of $O(\hbar^2)$ 
arise in the Hamiltonian which are neglected in the current analysis.
\end{itemize}

\subsection{The transition state part}\label{tssol}
In the transverse  degrees of freedom corresponding to  $(\qhat,\phat)$, 
we suppose a discrete spectrum parametrised by $\I$ in the following 
way
\[
H_\TS(\qhat,\phat,\I)\ket{\varphi_k(\I)} 
                       = E_\TS^k(\I)\ket{\varphi_k(\I)}.
\]
Here we suppose that a partial symbol $H_\TS(\qhat,\phat,\I)$
is defined by replacing $\Ih$ by its eigenvalue $\I$ in 
$H_\TS(\qhat,\phat,\Ih)$. If $H_\TS(\qhat,\phat,\Ih)$ is given
as a series of the form (\ref{expandH}) then we can write, concretely,
\[
H_\TS(\qhat,\phat,\I) = H_0(\qhat,\phat) 
+ \I H_1(\qhat,\phat) + \I^2 H_2(\qhat,\phat) +\cdots.
\]
In many interesting chemical applications, 
problems arise which have a Morse or Van der Waals 
type potential in the transverse degree of freedom for which the 
spectrum  of $H_\TS(\qhat,\phat,\I)$ is discrete at the bottom but 
becomes continuous above a threshold. We will confine ourselves, however, 
to energies at which asymptotically propagating scattering states correspond to
reactant molecules in bound states, and for these cases the
continuous spectrum (of $\hat{H}_\TS$) does not participate. Rather 
than adjusting notation here to incorporate the continuous part of the 
spectrum we simply suppress it notationally and consider 
states labelled by the discrete index $k$ only.

Results like those in \cite{me} can be obtained by ignoring the 
$\I$-dependence of the states $\ket{\varphi_k(\I)}$ and approximating 
them by  $\ket{\varphi_k(0)}$ (or equivalently keeping only the leading part 
$H_0(\qhat,\phat)$ in the expansion above), but here we want to
investigate the effect of coupling between the reaction and transverse 
degrees of freedom seen in the full Hamiltonian.
For a fixed $\I$ we can assume that these states form an orthonormal 
set
but note that we should assume in general that
\begin{equation}\label{nonON}
\braket{\varphi_{k'}(\I')}{\varphi_k(\I)} \neq \delta_{kk'}
\end{equation}
if $\I\neq\I'$. This is the main point complicating the
following analysis and means that we should be wary of assuming 
``obvious'' results when describing issues of normalisation.

We will now describe how the discrete eigensolutions of the 
transverse problem combine with the scattering solutions
found in the reaction-coordinate degree of freedom.

\subsection{Eigenfunctions of the total Hamiltonian}
With the conventions described in Sections~\ref{rcsol} and \ref{tssol},
\[
\ket{\Psi_{\I,k}} = \psi_\I(x)\ket{\varphi_k(\I)}
\]
are eigenstates of the full Hamiltonian $\Hh$ in mixed position-bra-ket 
notation. These solutions satisfy
\[
\Hh\ket{\Psi_{\I,k}}= E_k(\I)\ket{\Psi_{\I,k}},
\]
where 
\[
 E_k(\I) = E_\RC(\I) + E_\TS^k(\I).
\]
Note that the states $\ket{\Psi_{\I,k}}$  are separable in form
even though the Hamiltonian itself is not strictly speaking
separable, as discussed at the beginning of this section. This 
nonseparability manifests itself through the dependence 
of the transverse 
eigenstates $\ket{\varphi_k(\I)}$ on $\I$ and in particular through 
the nonorthogonality condition (\ref{nonON}).

Our aim is eventually to express results in such a way that
explicit reference to the normal form transformation is unnecessary.
For this purpose it is preferable to label states with the total 
energy instead of the reaction action, since the energy is defined
independently of the representation used.
For each value $E$ of the total energy let, $\I_k(E)$ be defined 
implicitly as a solution of
\[
E = E_k(\I).
\]
Although we cannot write explicit expressions for $\I_k(E)$,
we can suppose that in an energy range around threshold 
(where $E_\RC(\I) = -\lambda\I+O(\I^2)$), these functions
are well defined and single-valued for each $k$. We then define
scattering states labelled by the total energy $E$ and the mode 
number $k$ as follows. Let
\[
\ket{\Psi_{E,k}} =  \psi_{\I_k(E)}(x)\ket{\varphi_k(E)},
\]
where for short we write
\[
\ket{\varphi_k(E)} = \ket{\varphi_k((\I_k(E))}.
\]
Note that in view of (\ref{nonON}) we have
\begin{equation}\label{nonONE}
\braket{\varphi_{k'}(E)}{\varphi_k(E)} \neq \delta_{kk'}
\end{equation}
since on changing $k$ the action eigenvalue $\I=\I_k(E)$ 
changes.

It is possible to normalise these states so that
\[
\braket{\Psi_{E',k'}}{\Psi_{E,k}} = \delta(E-E')\delta_{kk'}
\]
in the usual way, but this convention turns out not to be particularly
useful for our purposes and we will not apply it. Instead we will
normalise these states so that they have unit flux in the incoming reaction
channel. A discussion of normalisation by flux will also be necessary
to appreciate how arbitrary scattering states may be constructed
from these separated solutions, so we will defer further consideration
of such issues until a method of flux calculation has been outlined in the 
next section.

\section{Fluxes and sectional inner products}\label{Fluxsec}
Since the normal-form Hamiltonian is not of kinetic-plus-potential
type, we cannot use the usual definition of current
\[
\J = \frac{\hbar}{2im}\left(\Psi^*\nabla\Psi-\Psi\nabla\Psi^*\right)
\]
to calculate fluxes. There is a simple generalisation, however, which 
works for arbitrary Hamiltonians.

Let $\hat{\Theta}$ be a Hermitian 
operator which projects to one side of a section $\Sigma$ which has 
codimension one in phase space. 
For example, if $\Sigma$ is defined by fixing a 
configuration space coordinate then in position representation
$\hat{\Theta}$ can represent 
multiplication by the characteristic function of a region which  
has boundary $\Sigma$. More generally, $\hat{\Theta}$ can be 
an operator for which a classical symbol such as the Weyl symbol rises 
from zero to unity in a classically small strip around $\Sigma$.
Then the flux of a state $\ket{\Psi}$ across $\Sigma$ is
\begin{equation}\label{genflux}
F = \langle\Psi\dbar_\Sigma\Psi\rangle,
\end{equation}
where we define a sectional overlap by
\begin{equation}\label{sectionaloverlap}
\langle\Phi\dbar_\Sigma\Psi\rangle = \frac{1}{i\hbar}
\braopket{\Phi}{[\hat{\Theta},\Hh]}{\Psi}.
\end{equation}
The notation here is adapted from \cite{APCW} although a factor of 
$i\hbar$ has been introduced which will simplify matters later.
Flux defined in this way is an integral part of transition-state
calculations in the chemical literature (see \cite{MillerFar,MillerQTST} 
for example) and similar ideas are used in \cite{CdVP3,AK}.
It is easily verified that if  $\hat{\Theta}$ represents multiplication
by the characteristic function of a region in configuration space
then the flux reduces to the standard case of a surface integral 
of the current $\J$  over the boundary.
Although a flux calculation requires only the diagonal case in 
(\ref{sectionaloverlap}), it is useful to allow the nondiagonal 
case in the definition of sectional overlap. We will find in particular
that the space of scattering solutions with a given total energy $E$
can be identified with a quantised surface of section
in one of the channels and the sectional overlap then provides a natural 
inner product for the corresponding Hilbert space.

\subsection{Fluxes for the separated scattering states}
We will now apply this generalised construction to compute fluxes in 
the normal form representation. Let the section $\Sigma$ be chosen so 
that $\hat{\Theta}$ can be constructed in terms of the $(\Qh,\Ph)$ 
operators alone and commutes with  $\qhat$ and $\phat$. In phase space 
this means
that $\hat{\Theta}$ projects onto a region of phase space defined by
a subset of the $QP$ plane and independent of the $(\q,\p)$ coordinates. 
Such a choice is natural if we view the complete system as a pinched 
waveguide (Figure~\ref{wgfig}) in which a section
obtained by fixing reaction coordinates is used to define fluxes in 
and out of the reactant and product channels.

\begin{figure}[h]
\vspace*{0.0cm}\hspace*{4.5cm}
\psfig{figure=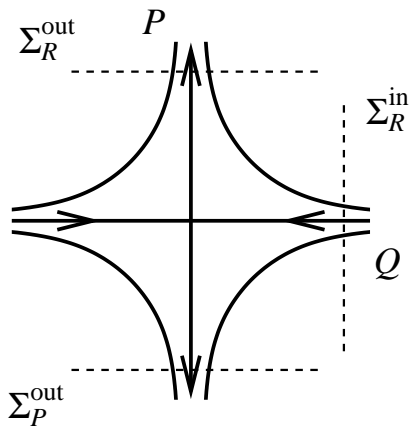,height=2.5in}
\vspace*{0.2in}
\caption{Sections which measure in and out fluxes in the
product and reactant channels are illustrated schematically. Sections
in full phase space are defined by fixing either the coordinate $Q$ or
the coordinate $P$ appropriately in each case.}
\label{QPsecfig}
\end{figure}

With the generalised definition of flux in (\ref{genflux}) 
we are not, however, confined to fluxes across surfaces in 
configuration space and are free to define sections
in phase space which distinguish incoming from outgoing flux in the 
reactant and product channels. For example, referring to 
Figure~\ref{QPsecfig}, 
a section $\Sigma_R^\in$ defined by a vertical line in the right-half 
of the $QP$-plane measures incoming flux in the reactant channel, whereas 
a horizontal line  $\Sigma_R^\out$ in the upper half plane measures
outgoing flux in the reactant channel and a horizontal line
$\Sigma_P^\out$ in the lower half plane measures outgoing flux in the
product channel. Fluxes across sections such as these corresponding to 
horizontal and vertical lines in the $QP$ plane are especially easily 
computed in the $QP$ representation. Details are given in \ref{1d}
for the  one-dimensional scattering solutions described in 
section~\ref{rcsol} --- here we simply note that relative
fluxes are found to be of the form given in (\ref{TR}). Fluxes in 
the full system can be understood on 
the basis of these one-dimensional calculations in the following way.

By an abuse of notation, let us denote a one-dimensional flux
for one of these reaction-coordinate sections by
\begin{equation}\label{Iflux}
\langle\psi_\I\dbarr_\Sigma\psi_\I\rangle 
= -\frac{1}{i\hbar}\braopket{\psi_\I}{[\hat{\Theta},\Ih]}{\psi_\I}.
\end{equation}
Although we use similar notation to (\ref{genflux}), it should be 
emphasised that this one-dimensional
flux differs in having $\Ih$ and not a Hamiltonian in the commutator
and is not therefore a straightforward physical flux. There is also a 
minus sign, which compensates for the fact that $I(Q,P)$ generates
a flow that is opposite in direction to the physical flow generated by 
the Hamiltonian --- corresponding to $E_k'(\I)<0$ in the discussion below. 
Although distinct from the physical flux, this quantity
is prominent in the physical answer and the notation is useful 
for that reason. 

We will see that in the full space the separated scattering states
$\ket{\Psi_{\I,k}}$ have sectional overlaps of the form
\begin{equation}\label{Influx}
\langle\Psi_{\I,k'}\dbar_\Sigma\Psi_{\I,k}\rangle 
= -E_k'(\I)\langle\psi_\I\dbarr_\Sigma\psi_\I\rangle \delta_{kk'}.
\end{equation}
Alternatively, using states labelled by the total energy
and normalised so that
\begin{equation}\label{normE}
\ket{\tilde{\Psi}_{E,k}} = \sqrt{-\I_k'(E)}\; \ket{\Psi_{\I_k(E),k}},
\end{equation}
we have
\begin{equation}\label{normfluxE}
\langle\tilde{\Psi}_{E,k'}\dbar_\Sigma\tilde{\Psi}_{E,k}\rangle 
= \langle\psi_{\I_k}\dbarr_\Sigma\psi_{\I_k}\rangle \delta_{kk'}.
\end{equation}
The main point here is that the multidimensional scattering
states have fluxes, and therefore probabilities of reflection
and transmission, which reduce exactly to the one-dimensional
case and we can apply (\ref{TR}) to scattering states of an 
arbitrary Hamiltonian in normal form. The transmission and reflection
probabilities of the states $\ket{\tilde{\Psi}_{E,k}}$ can therefore
be written
\begin{equation}\label{TRn}
T_k(E) 
=\frac{1}{1+\e^{2\pi\I_k(E)/\hbar}}
\quad\mbox{and}\quad R_k(E) = \frac{1}{1+\e^{-2\pi\I_k(E)/\hbar}}
\end{equation}
and we can now use this as a basis with which to treat reactivity
of an arbitrary stationary state.

We should emphasise that in view
of the nonorthogonality of the transverse modes as expressed in 
(\ref{nonON}) and (\ref{nonONE}), and in contrast to separable problems, 
these identities extending the 
one-dimensional results are not at all 
obvious. To prove them we consider separately the diagonal and 
nondiagonal cases, and treat them as follows.

\vspace{24pt}
\noindent {\bf Derivation of (\ref{Influx}) in the diagonal case}\\
We treat the diagonal case $k=k'$ first. Assume that $\Hh_\TS$ can 
be expanded in the form (\ref{expandH}) and use 
\[
[\hat{\Theta},\Ih^m] = [\hat{\Theta},\Ih]\Ih^{m-1} 
+  \Ih[\hat{\Theta},\Ih]\Ih^{m-2} + \cdots + \Ih^{m-1}[\hat{\Theta},\Ih]
\]
to deduce that
\[
\braopket{\psi_{\I}}{ [\hat{\Theta},\Ih^m]}{\psi_{\I}} 
= m\I^{m-1}\braopket{\psi_\I}{[\hat{\Theta},\Ih]}{\psi_\I}
\]
and therefore that
\begin{eqnarray*}
\braopket{\Psi_{\I,k}}{ [\hat{\Theta},\Hh_\TS]}{\Psi_{\I,k}}
&=&\sum_m m\I^{m-1} \braopket{\psi_\I}{[\hat{\Theta},\Ih]}{\psi_\I}
\braopket{\varphi_k(\I)}{H_m(\qhat,\phat)}{\varphi_k(\I)}\\[6pt]
 &=& \braopket{\psi_\I}{[\hat{\Theta},\Ih]}{\psi_\I}
\braopket{\varphi_k(\I)}
{\dydxv{\Hh_\TS(\qhat,\phat,\I)}{\I}} {\varphi_k(\I)}\\[6pt]
&=& \braopket{\psi_\I}{[\hat{\Theta},\Ih]}{\psi_\I}
\dydxv{E_\TS^k(\I)}{\I},
\end{eqnarray*}
where in the last line we have invoked the Feynman-Hellman theorem.
On adding a similar calculation for $\Hh_\RC$ we get the claimed result.
The important feature is that the factor $E_k'(\I)$ does not 
depend the choice of section across which to measure flux and 
relative fluxes reduce to the one-dimensional case.

\vspace{24pt}
\noindent {\bf Derivation of (\ref{Influx}) in  the nondiagonal case}\\
To treat the nondiagonal case we use the identity
\begin{eqnarray*}
\braopket{\psi_{\I'}}
{[\hat{\Theta},\Ih^m]}{\psi_{\I}}
&=&\braopket{\psi_{\I'}}
{\Bigl([\hat{\Theta},\Ih]\Ih^{m-1}+\Ih[\hat{\Theta},\Ih]\Ih^{m-2}
+\cdots\Ih^{m-1}[\hat{\Theta},\Ih]\Bigr)}
{\psi_\I}\\[6pt]
&=& [\I^{m-1} + \I'\I^{m-2}+\cdots \I'^{m-1}] 
\braopket{\psi_\I'}{[\hat{\Theta},\Ih]}{\psi_\I} \\[6pt]
&=& \frac{\I^m-\I'^m}{\I-\I'}
\braopket{\psi_\I'}{[\hat{\Theta},\Ih]}{\psi_\I}
\end{eqnarray*}
to deduce that
\[
\braopket{\Psi_{\I',k'}}
{[\hat{\Theta},\Ih^m\Hh_m]}{\Psi_{\I,k}}
= \braopket{\psi_{\I'}}{[\hat{\Theta},\Ih]}{\psi_\I}
\braopket{\varphi_{k'}(\I')}{\left[
\frac{\I^m\Hh_m-\I'^m\Hh_m}{\I-\I'}
\right]}{\varphi_k(\I)}.
\]
On summing over $m$ and doing a similar calculation
for the reaction-coordinate part of the Hamiltonian
we find that
\[
\langle{\Psi}_{\I',k'}\dbar_\Sigma{\Psi}_{\I,k}\rangle 
=-\frac{E_k(\I)-E_{k'}(\I')}{\I-\I'}
\langle\psi_{\I'}\dbarr_\Sigma\psi_{\I}\rangle
\braket{\varphi_{k'}(\I')}{\varphi_k(\I)}.
\]
In terms of the energy-labelled states
$\ket{{\Psi}_{E,k}} = \ket{\Psi_{\I_k(E),k}}$,
we have
\[
\brasket{\Psi_{E',k'}}{\Sigma}{\Psi_{E,k}}
=-\frac{E-E'}{\I_k(E)-\I_{k'}(E')}
\langle\psi_{\I_{k'}(E')}\dbarr_\Sigma\psi_{\I_k(E)}\rangle
\braket{\varphi_{k'}(E')}{\varphi_k(E)}.
\]
These sectional overlaps vanish as $E'$ approaches  
$E$ unless $k=k'$, in which case we recover the 
diagonal result. This completes the derivation of 
(\ref{Influx}).

\subsection{Fluxes for general scattering states and reduced Hilbert space}
Let $\H_R^\in$ denote the subspace of scattering states of a fixed total 
energy $E$ which are incoming on the reactant side and have no incoming 
flux on the product side. In the applications we consider the scattering 
problem will asymptote to a waveguide-type problem in which there are 
a finite number $\Nchan$ of propagating channels (corresponding to the 
energetically accessible bound states of the reacting molecules) and 
the space $\H_R^\in$ will therefore be finite-dimensional. We assert that
the space $\H_R^\in$ is in fact essentially a quantisation of a classical 
surface of section (in the incoming reactant channel) in the sense of 
Bogomolny \cite{Bog}, with the inner product being given by sectional 
overlaps of the form in (\ref{sectionaloverlap}).

Let us write a general asymptotically-propagating scattering state
in the form
\begin{equation}\label{defPhi}
\ket{\Phi} = \sum_{k=1}^{\Nchan} a_k \frac{\ket{\tilde{\Psi}_{E,k}}}
{\sqrt{\langle\psi_{\I_k}\dbarr_{\Sigma_R^\in}\psi_{\I_k}\rangle}}.
\end{equation}
Then the total incoming flux for this state is
\[
F_R^\in = \langle{\Phi}\dbar_{\Sigma_R^\in}{\Phi}\rangle =  
\sum_{k=1}^{\Nchan}|a_k|^2
\]
while in view of (\ref{normfluxE}) and (\ref{TRn}) the outgoing 
flux in the reactant channel is
\[
F_R^\out = \langle{\Phi}\dbar_{\Sigma_R^\out}{\Phi}\rangle =
 \sum_{k=1}^{\Nchan} \frac{1}{1+\e^{2\pi\I_k(E)/\hbar}} |a_k|^2
\]
and the outgoing flux in the product channel is
\[
F_P^\out = \langle{\Phi}\dbar_{\Sigma_P^\out}{\Phi}\rangle =
 \sum_{k=1}^{\Nchan} \frac{1}{1+\e^{-2\pi\I_k(E)/\hbar}} |a_k|^2.
\]
Furthermore, between any two such scattering states the sectional
overlap
\[
\langle{\Phi'}\dbar_{\Sigma_R^\in}{\Phi}\rangle =\sum_{k=1}^{\Nchan} a_k'^*a_k
\]
provides a natural inner product for $\H_R^\in$.

We can regard $\H_R^\in$ abstractly as a space spanned by an
orthonormal basis $\{\ket{k}\}_{k=1}^{\Nchan}$ whose elements 
\[
\ket{k}\sim \frac{\ket{\tilde{\Psi}_{E,k}}}
{\sqrt{\langle\psi_{\I_k}\dbarr_{\Sigma_R}\psi_{\I_k}\rangle}}
\] 
are in one-to-one correspondence with the scattering states
$\ket{\tilde{\Psi}_{E,k}}$, normalised to have unit incoming flux. 
A general element 
\[
\ket{\varphi}=\sum_{k=1}^{\Nchan}  a_k \ket{k}
\]
of this reduced space can be extended to a scattering state 
of the form given in (\ref{defPhi}) for which the total energy is 
fixed but which is not necessarily an eigenstate of the transverse 
Hamiltonian $\Hh_\TS$.
The inner product between two reduced states can be defined through
sectional overlaps
\[
\braket{\varphi'}{\varphi} = 
\langle{\Phi'}\dbar_{\Sigma_R}{\Phi}\rangle
\]
of the corresponding extended states.

The benefit of this abstraction is that we can compute outgoing
fluxes in the product channel using matrix elements
\begin{equation}\label{matel}
F_P^\out = \braopket{\varphi}{\R(E)}{\varphi}
\end{equation}
of a reaction operator
\begin{equation}\label{intR}
\R(E) = \sum_{k=1}^{\Nchan} \frac{\ket{k}\bra{k}}
{1+\e^{2\pi\I_k(E)/\hbar}}
= \sum_{k=1}^{\Nchan} \frac{\e^{-2\pi\I_k(E)/\hbar}}
{1+\e^{-2\pi\I_k(E)/\hbar}}\ket{k}\bra{k}
\end{equation}
which is diagonal in this basis. By writing the outgoing flux
as a matrix element of an operator defined on $\H_R^\in$ we have
in large part achieved the goal of this paper, which is to 
generalise the construction in \cite{me} so that there is
no longer a restriction to states supported in a classically 
small region of phase space. In fact, the only restriction
on incoming states here is that they should be supported in the 
region of phase space where the normal form in (\ref{cnf}) provides
an accurate description of dynamics. Although
undoubtedly a restricted subset of phase space, this domain
has classical dimensions.

The current version is tied to the normal form, however and in order
for this operator to be of any practical use, 
we really need a way of constructing it which does not call 
on us explicitly to construct the normal form or the basis vectors 
$\ket{k}$. As a start in this direction, note that if we denote 
\begin{equation}\label{primT}
\T(E) = \sum_k \e^{-2\pi\I_k(E)/\hbar}\ket{k}\bra{k},
\end{equation}
then $\R(E)$ can be written in the form (\ref{uniR}) promised in 
the introduction.
This is precisely the form given in \cite{me}, where
$\T(E)$ was a {\it tunnelling operator} defined as a quantised 
surface of section map in the neighbourhood of a complex periodic 
orbit. We will show in the Section~\ref{nonnfsec} that the same interpretation 
can be imposed on $\T(E)$ in the present case with the difference that,
unlike in \cite{me}, restriction to a neighbourhood of the periodic 
orbit small enough for linearised dynamics to be used is no longer 
necessary.

\subsection{The microcanonical cumulative reaction probability}
The trace
\begin{equation}\label{micro}
N(E) = \Tr\, \R(E) =  \sum_{k=1}^M\frac{1}{1+\e^{2\pi\I_k(E)/\hbar}}
\end{equation}
of the reaction operator is the so-called microcanonical 
cumulative reaction probability. Results for $N(E)$ equivalent
to those that would be obtained by using linearised dynamics 
in $\R(E)$ in the manner of \cite{me} were obtained by Miller in 
\cite{Millermicro}. Semiclassical approximations  for $N(E)$
have also been given in \cite{MillerFar,MillerHern1} which 
include nonlinear effects in the transition state degrees of freedom 
by using expansions in the transition-state quantum numbers (a 
related treatment of tunnelling using normal-form coordinates
has been given in \cite{Sri}). A thermalised
version has been given in \cite{MillerHern2} and see
\cite{Millerlatest} for a semiquantum calculation.
A discussion emphasising the fluctuations that occur in $N(E)$ as
the summands in (\ref{micro}) switch on with increasing $E$
can be found in \cite{Cha1,Cha2}. We also remark that a discussion 
of $N(E)$ has recently been given in \cite{HCN} which uses the same 
language of normal forms that we use here.

The major benefit of the current work is that, once a prescription has 
been given in the next section for computing $\R(E)$ without recourse
to the normal form, we will have an explicit prescription
for distributing the total reaction probability in $N(E)$ 
among incoming states or, equivalently, using the Wigner-Weyl formalism,
for understanding how the reaction probability is distributed in
phase space. In principle, this calculation can be implemented simply
by computing complex trajectories near the complex periodic orbit
used in \cite{me} and does not require a particular deconstruction of the 
Hamiltonian such as provided by a normal form (although normal forms 
do help in inverting the operator $1+\T(E)$ in (\ref{uniR}) as we will 
discuss).

\section{Getting away from the normal form}\label{nonnfsec}

Equation~(\ref{uniR}) promises the ability to treat
reaction problems without having to deal explicitly with the
normal form and separated stationary states. In this section 
we show how this can be done by interpreting $\T(E)$ as a
tunnelling operator in the sense of \cite{me,APCW} which can be understood
independently of the normal form construction.

\subsection{The complex return map in normal form coordinates}\label{thetasec}
We first describe how a complex return map is expressed
in terms of normal form coordinates. Let $(I,\theta)$ be action
angle variables in the $QP$ plane such that in the positive quadrant
\begin{eqnarray*}
Q &=& \sqrt{I}\e^\theta\\
P &=& \sqrt{I}\e^{-\theta}.
\end{eqnarray*}
We have $I = QP$, consistent with previous notation. Although
the system is hyperbolic and not normally associated with periodic 
motion, there is in fact an imaginary period, expressed by
the identities
\begin{eqnarray*}
Q(\theta+2\pi i) &=& Q(\theta)\\
P(\theta+2\pi i) &=& P(\theta).
\end{eqnarray*}
In other words, the Hamiltonian flow generated by $I$ in the complexified
$QP$ plane has period $2\pi i$. We will now show that an extension of
this periodic flow to the full system can be used to define
a complex Poincar\'e return map and that this map has a fixed point
corresponding to a complex periodic orbit.

Let us restrict initial conditions to a surface of section $\Sigma_R^\in$
defined by fixing $Q$ as in Figure~\ref{QPsecfig}, along with the total
energy $E$, and let the condition
\begin{equation}\label{implicitdef}
E 
= H(\q,\p,I)
\end{equation}
implicitly define the function  $h(\q,\p,E)$ on $\Sigma_R^\in$ by
\[
I = h(\q,\p,E)-e(E).
\]
Here we regard $(\q,\p)$ as canonical coordinates for $\Sigma_R^\in$
and $e(E)$ is defined so that the minimum of  $h(\q,\p,E)$ on 
$\Sigma_R^\in$ is zero, as described more explicitly below. Differentiating 
(\ref{implicitdef}) with respect to the transverse coordinates $\q$ and 
$\p$ while keeping $E$ fixed gives
\[
0 = \dydxv{H}{I}\nabla h + \nabla H,
\]
which in turn gives
\[
X_h = -\frac{1}{\dot{\theta}}\tilde{X}_H 
= -\spinor{\ds\dydxvo{\q}{\theta}}{\ds\dydxvo{\p}{\theta}},
\]
where $\tilde{X}_H$ denotes the projection of the full flow vector 
$X_H$ defined by $H$ onto the transverse degrees of freedom 
$(\q,\p)$ and we have used $\dot{\theta}=\dydxh{H}{I}$. The flow
defined by $X_h$ can therefore be regarded as a restriction to
the $(\q,\p)$ degrees of freedom of the full flow, reparametrised
so that time $t$ is replaced with the angle variable $\theta$.

Letting $\theta$ evolve from $0$ to the final value $2\pi i$, trajectories
are described in full phase space which start on $\Sigma_R^\in$
and return to it --- recall that the $(Q,P)$ coordinates which are
used to define $\Sigma_R^\in$ are periodic under this evolution.
Integrating the flow vector $-X_h$ for a time $2\pi i$ then generates 
a complex symplectic map 
\[
\F:\Sigma_R^\in\to\Sigma_R^\in
\]
which we can denote by
\begin{equation}\label{cmap}
\F = \exp[-2\pi iX_h]
\end{equation}
in Lie-algebraic notation. This is precisely the classical map used to 
construct the tunnelling operator in \cite{me}. 

Before describing explicitly how the quantisation works,
it is helpful to see how the complex periodic orbit 
which provides a fixed point of $\F$ arises in normal form coordinates.
By construction, the quadratic part of the Hamiltonian $H(\q,\p,I)$
is elliptic in the transverse degrees of freedom. As a result,
the sectional Hamiltonian $h(\q,\p,E)$ has a minimum 
$(\q_e(I),\p_e(I))$, for sufficiently small $I$ at least, for which
\[
\nabla h (\q_e(I),\p_e(I),E) = 0
\]
and which coincides with the origin of  $\Sigma_R^\in$ in the threshold 
case $I=0$. We define $e(E)$ above so that $h(\q_e,\p_e,E)=0$ and a 
Taylor expansion of $h(\q,\p,E)$ about $(\q_e(I),\p_e(I))$ begins 
with quadratic terms. This minimum is an equilibrium of the flow 
defined by $X_h$ on $\Sigma_R^\in$ and is therefore a fixed point of $\F$. 
In full phase space, the trajectory starting with coordinates 
$(\q_e(I),\p_e(I))$ on  $\Sigma_R^\in$ evolves so that the coordinates 
$I$ and $(\q,\p)$ are fixed and defines a periodic orbit as $\theta$ 
evolves from $0$ to $2\pi i$. The time period corresponding to this 
evolution is $-i\tau(E)$ where
\[
\tau = -\frac{2\pi}{\dot{\theta}} 
= -\frac{2\pi}{\dydxh{H(\q_e,\p_e,I)}{I}} 
= 2\pi e'(E).
\]
and its action is an imaginary number $S(E)=iK_0(E)$ where
\[
K_0 = \frac{1}{i}\oint I\d\theta = 2\pi I = -2\pi e(E).
\]
In \cite{me} a linearisation of dynamics about this complex periodic
orbit was used to approximate $\R(E)$ which here corresponds to 
truncating a Taylor series of $h(\q,\p,E)$ about $(\q_e(I),\p_e(I))$ at 
quadratic order. The essential conclusion of this paper is that a 
complete description of the reaction operator can be achieved simply 
by replacing this truncation with the full sectional Hamiltonian 
$h (\q,\p,E)$.

\subsection{The tunnelling operator}\label{Tsec}
The tunnelling operator is defined to be a quantisation of the classical
map $\F$ \cite{me,APCW}. Using (\ref{cmap}) we can write concretely, 
\[
\T = \e^{2\pi (e-\hh)/\hbar},
\]
where $\hh-e$ is the restriction of the operator $\Ih$ to
the quantum analogue $\H_R^\in$ of $\Sigma_R^\in$ defined
in the Section~\ref{Fluxsec}.  In that case
\[
(\hh-e)\ket{k} = \I_k(E)\ket{k}
\]
and
\[
 \e^{2\pi (e-\hh)/\hbar}\ket{k} =  \e^{-2\pi \I_k(E)/\hbar}\ket{k}
\]
and the identification in the previous section  of $\T(E)$ in 
(\ref{primT}) as a tunnelling operator is confirmed.

It should be emphasised that while the sectional Hamiltonian $h$
was used in making this identification, it is not necessary to
construct it explicitly, or even to refer to it, in order
to construct the map $\F$ and to approximate its quantum analogue 
$\T(E)$ semiclassically. The map  $\F$ can be constructed simply
by integrating orbits as described in references \cite{me,APCW}.
From the dynamical characteristics of these orbits, Van Vleck-type
approximations for $\T(E)$ can be written as described in \cite{Bog}, 
for example. Alternatively, the Weyl symbol of $\T(E)$ can be obtained 
as described in \cite{Berryscar}. Coherent state representations
are also possible (\cite{Adachi,Bar} and references therein). In all of 
these approaches
the dynamical information needed is naturally provided as a result 
of the orbit computation and $h(\q,\p)$ is not needed explicitly.

We note finally that while it would be quite easy to write
a version of (\ref{intR}) that describes the full scattering matrix 
in the normal form representation, so that we could determine
the distribution of product states for each reactant state,
it is less obvious how the normal form result could be interpreted 
in a basis-independent way in that case. Any such reformulation
would have to take into account the fact that orbits contributing 
to the scattering matrix itself \cite{MillerSmatrix} depend singularly 
on initial conditions near the reacting boundary. The simplicity of the 
normal form representation suggests, however that such a uniformisation 
might be feasible, although we do not pursue it here.

\subsection{Asymptotic basis for the reaction operator}
Although the reaction operator is diagonal with respect
to the stationary states $\ket{\tilde{\Psi}_{E,k}}$
computed in terms of the normal form, it will not in general 
be diagonal in the basis of asymptotically decoupled stationary 
states that is used to write the scattering matrix. In fact,
since the normal form will in general only provide an accurate
description of dynamics in a neighbourhood of the transition 
state, we need an independent method to describe how these 
states can be extended to the asymptotic regions of the 
reactant channel and the operator $\R(E)$ written in the standard 
asymptotic basis. 

To achieve this we note simply that once the transmission problem has 
been solved locally in the transition state region, the solution 
can extended to the asymptotic region by applying
quantised surface-of-section maps, such as described in \cite{Bog}, 
for example. Once outside the transition state region, these mappings 
can be constructed on the basis of primitive semiclassical 
approximations and amount in the classical picture simply to 
conjugating the complex Poincar\'e map $\F$ with standard real ones.
As when transforming to the quantum normal form in the first place, 
the end result of this process is easily stated and the details
omitted in the interests of brevity since they follow discussions 
elsewhere \cite{Bog}.

To be more specific about this conjugation, let the coordinates
$(x,y,p_x,p_y)$ be as described in the introduction and let the 
Hamiltonian decouple asymptotically in the 
reactant channel according to
\begin{equation}\label{decouple}
H(x,y,p_x,p_y) \simeq H_\as(y,p_y) + H_\tr(x,p_x).
\end{equation}
Here $H_\as(y,p_y)$ describes the internal vibrational motion of the
reacting molecules and  $H_\tr(x,p_x)$ is a kinetic energy term
for the relative motion of centres of mass. In the simplest 
atom-diatom collinear case we might have
\[
H_\tr(x,p_x) = \frac{p_x^2}{2M}
\]
where $M$ is the atom-diatom  relative mass and
\[
 H_\as(y,p_y) =  \frac{p_y^2}{2m} + V_{\rm AB}(y)
\]
where $m$ is the diatomic relative mass and $V_{\rm AB}(q)$ the
diatomic interaction potential.

The separated asymptotic solutions are denoted
\[
\ket{\Phi_{n,E}} = \chi_{E,n}(x)\ket{\psi_n}
\]
where $\chi_{E,n}(x)$ is a plane-wave eigenfunction of
$\Hh_\tr$ with energy $E-E_n$ and normalised to have unit incoming flux 
while $\ket{\psi_n}$ is an eigenstate of internal dynamics with energy
$E_n$. There is a phase convention implicit in writing the scattering 
matrix, which in the present case amounts to specifying the phase of 
$\chi_{E,n}(x)$. In the standard plane-wave case
\[
\chi_{E,n}(x) = \frac{\e^{i k_n x}}{\sqrt{v_n}},
\]
where $v_n=\hbar k_n$ ensures unit incoming flux, we can understand
the phase of $\chi_{E,n}(x)$ as being fixed at $x=0$. 

\begin{figure}[h]
\vspace*{0.2cm}\hspace*{3.5cm}
\psfig{figure=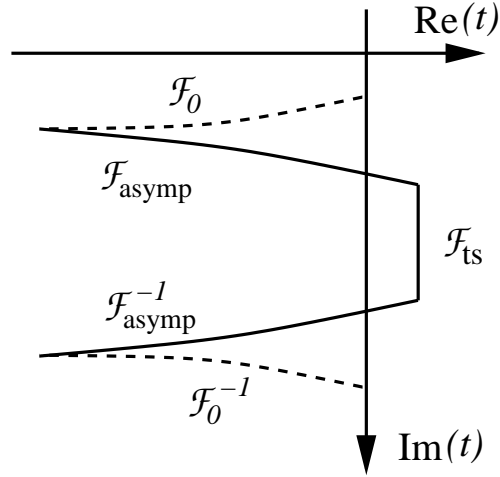,height=2.5in}
\vspace*{0.2in}
\caption{The path in the complex time plane which achieves
the conjugation (\ref{conj}) is illustrated schematically.
The dashed lines indicate evolution under decoupled dynamics.
For the complex periodic orbit defining a fixed point of $\F$,
the contour segments corresponding to $\F_0$, $\F_\as$ and their inverses 
are parallel to the real axis, while the segment for $\F_\TS$ is parallel to 
the imaginary axis. The time evolution of each segment develops both 
real and imaginary parts as the initial condition moves away from this 
fixed point.}
\label{tfig}
\end{figure}

The phase convention in the asymptotic regime is then obtained 
by extending the complete scattering state $\ket{\Phi_{n,E}}$ from a section
$\Sigma_R^0$ defined by $x=0$ to a section $\Sigma_R^\as$ defined by a 
large and negative value of $x$. This extension is affected in the 
semiclassical scheme \cite{Bog} by quantising a surface of section mapping
\[
\F_0:\Sigma_R^0\to \Sigma_R^\as
\]
constructed using the uncoupled dynamics of the Hamiltonian in 
(\ref{decouple}) --- since the plane wave is travelling to the right, 
this will be achieved in negative time.
Scattering of the resulting incoming state then proceeds by first 
mapping from the section $\Sigma_R^\as$ to a section $\Sigma_R^\TS$ 
near the transition state using the fully coupled dynamics,
\[
\F_\as: \Sigma_R^\as\to \Sigma_R^\TS.
\]
If $\Sigma_R^\TS$ is in the domain of the normal form then its
transmission probability is understood using a complex return map 
\[
\F_\TS: \Sigma_R^\TS\to \Sigma_R^\TS
\]
of the kind described in Section~\ref{thetasec}. Finally, we complete
the transformation to the asymptotic basis by mapping 
back to the asymptotic section $\Sigma_R^\as$ using the inverse 
$\F_\as^{-1}$ of $\F_\as$ and then mapping to $\Sigma_R^0$ using 
the inverse $\F_0^{-1}$. The end result is a map
\begin{equation}\label{conj}
\F = \F_0^{-1}\F_\as^{-1}\F_\TS\F_\as\F_0
\end{equation}
which is conjugate to $\F_\TS$ but which is adapted to the phase convention
used for the scattering matrix. A quantisation of this map gives a 
tunnelling operator $\T(E)$ appropriate to the asymptotic basis
defined by the states  $\ket{\Phi_{n,E}}$.

Note that by conjugating $\F_\TS$ by $\F_\as$, we can extend the
return map far beyond the domain where the normal form applies. The
outer conjugation by $\F_0$ is not strictly necessary to calculate 
reaction probabilities of incoming states $\ket{\Phi_{n,E}}$
since the transverse parts $\ket{\psi_n}$ are eigenfunctions
of the quantisation of  $\F_0$ and the resulting eigenphases
cancel in the transformed version
\[
F_P^\out = \braopket{\psi_n}{\R(E)}{\psi_n}
\]
of (\ref{matel}). However phases are important for cross 
terms if we want to treat general scattering states
\[
\ket{\Phi} = \sum_{n=1}^M c_n \ket{\Phi_{n,E}}
\]
and they are also important for representations of $\R(E)$ in 
phase space. We also remark that $\F_\as$ will not in general commute
with $\F_\TS$ and the reaction operator will therefore {\it not} be 
diagonal in the asymptotic basis defined by the states 
$\ket{\Phi_{n,E}}$. On noting that the conjugation in (\ref{conj})
amounts simply to a change of representation so that $\Sigma_R^0$
can be identified with the section $\Sigma_R^\in$, however, 
we see that the tunnelling operator written in this basis is not
fundamentally different from the one described in Section~\ref{Tsec}.

Finally, we note that the conjugation in (\ref{conj}) is particular
to a waveguide problem in which the plane wave part is written
in terms of a coordinate $x$ with phases fixed at $x=0$. If different
conventions are used for the asymptotic coordinates or for the part
$\chi_{E,n}(x)$ of the scattering state, 
then the part $\F_0$ of the conjugation must be redefined accordingly.

\subsection{Limitations of the derivation}
No approximations are made in getting to the reaction-operator
form of the outgoing flux in (\ref{matel}) and (\ref{intR})
once the quantum normal form of the Hamiltonian in (\ref{qnf}) is written 
down. The sources of error are in transforming to the quantum normal form 
in the first place and in transforming to the asymptotic scattering basis
after the reaction operator has been found in the normal form representation
(\ref{matel}) and (\ref{intR}). We now comment on some of the issues
affecting this approach. 

The classical normal form is a formal series which describes
the dynamics locally in a neighbourhood of the transition state region.
It is clear that writing a quantum version of it as we do here is
a formal step which will ultimately require a more careful justification.
Issues of convergence become especially important if we pursue
the limit $\hbar\to 0$ in its literal sense and we have not addressed 
such questions here.  From a purely practical point of view, 
however, if one's aim is to achieve an approximation that works well
for a small but fixed value of $\hbar$, then it suffices to describe
the dynamics to a corresponding level of accuracy in phase space and
the normal form is certainly capable of that in the sorts of parameter 
regimes that arise in chemical applications \cite{us}.

Results of a numerical investigation are outlined in
\cite{us} which indicate that the expression in (\ref{uniR}) works very 
well when $\T(E)$ is computed directly from the complex orbits 
of the Poincar\'e return map $\F$. We also note that since (\ref{uniR})
can be interpreted theoretically without reference to the normal form, 
it seems natural to expect that it applies independently of the normal 
form itself. 
We therefore conjecture that, despite the limitations of the derivation 
presented here, Equation~(\ref{uniR}) is in fact ``classically exact'' in 
the sense that no errors arise from classical dynamics side of the calculation 
once $\T(E)$ is interpreted as the quantisation of $\F$
and the only approximation is the usual semiclassical one which 
vanishes as $\hbar\to 0$.

It is important to add the qualification, however, that even
if the conjecture is correct, there are good reasons to expect it 
to apply only locally, at least in the simple
form described in sections \ref{thetasec} and \ref{Tsec}. The return 
map $\F$ describes a unique image for initial conditions in a neighbourhood 
of the complex periodic orbit and for energies sufficiently close to threshold.
In practical terms, this means there is a unique complex solution 
satisfying the boundary conditions required of orbits by semiclassical 
approximation of $\T(E)$ in the usual representations \cite{me,us}. 
Sufficiently far away, however, bifurcations are likely to occur 
where this structure breaks down and these are not described by the 
current formulation. In fact, it has been shown in 
\cite{Takahashi} (and see \cite{Shudo,Onishi,Julia,Shudo2} for related 
work) that complex orbits contributing to the scattering matrix
can be chaotic and are subject to intricate pruning by the Stokes' phenomenon,
while numerical evidence \cite{us} suggests that the same
is true of the orbits contributing to $\T(E)$ sufficiently far from the
centre of the reacting region. 
At an even more basic level, the NHIM itself may undergo bifurcation
once the energy rises far enough above threshold and in this case
the whole bottleneck picture at the basis of our calculation is no 
longer correct. Global recrossing may occur and resonances arise
in the quantum mechanics which are not described by the simple picture
of transmission probability we have here. It should be stressed, however,
that whatever the limits are on the domain where contributing dynamics
are simple, they are independent of $\hbar$ and therefore have classical 
scales.

Finally, we remark that even though the tunnelling operator $\T(E)$
can be routinely approximated using semiclassical approximations,
the inverse of $1+\T(E)$ that occurs in (\ref{uniR}) is more 
problematic. Closed form analytic approximations are possible
if we know the sectional Hamiltonian $h(\q,\p)$ (see \cite{me} for 
the harmonic case) but more work is needed to provide
an approximation that works directly in terms of the map $\F$.

\section{Conclusion}\label{conclusion}
We have characterised the semiclassical transmission of 
waves across a phase-space bottleneck using a reaction operator constructed 
from a complex Poincar\'e mapping. In contrast to previous work 
\cite{me}, this construction is not restricted to energies and parts 
of phase space in a classically small neighbourhood of the
transition state at threshold.

A phase-space representation of the reaction operator will be 
largely supported in the classically reacting subset of phase space,
but will also incorporate tunnelling and other quantum effects
at the boundary of this region, where trajectories approach the
NHIM along its stable manifold. 
The only dynamical information needed to apply the approximation 
described in this paper is contained in the complex Poincar\'e 
mapping and explicit consideration of normal forms, or other
special assumptions regarding the dynamics such separability or
adiabatic approximation, are unnecessary. We also note that the 
particular trajectories used to define this map are well behaved at 
the reacting boundary.
Therefore, despite the fact that the fate 
of trajectories changes discontinuously across the reacting boundary, 
the  result here uniformly describes the transition from
classically allowed transmission inside the reacting region
to reaction entirely by tunnelling outside it.

We conclude by noting that in its current form the result here
is restricted to collinear problems. In order for the approach to be
used in completely realistic models, the marginally stable 
degrees of freedom associated with rotational symmetry
will need to be incorporated. This aspect needs further 
investigation.

\vspace{1cm}

\noindent {\bf Acknowledgements}\\
\noindent
This work was supported the European Network MASIE.

\appendix
\section{Scattering for the one-dimensional normal form}\label{1d}
The one-dimensional problem is especially easily solved in a 
representation in which the reaction action operator takes the form
(\ref{IQP}) and $\Ph$ acts on functions of $Q$ according to
\[
\Ph\psi(Q) = \frac{\hbar}{i}\psi'(Q).
\]
In this case the eigenvalue equation $\Ih\psi=\I\psi$ is a first order
differential equation
\begin{equation}\label{nf1d}
\frac{\hbar}{i}\left(Q\dydxvo{\ph{q}}{Q}+\ha\right)\psi(Q) = \I\psi(Q)
\end{equation}
and this can be solved in elementary terms without the complication of 
parabolic cylinder functions that arise in the conventional
representation of an inverted oscillator \cite{BM}.
This approach has in particular been exploited in \cite{CdVP1}
to treat scattering in one-dimensional networks of tori and we
refer to that publication for more detail of the 
following calculation. It is useful, however, to reiterate some of 
the main points here and to emphasise flux calculations, which form 
a basis for the discussion in the main text.

The eigenvalue equation (\ref{nf1d}) leads to doubly degenerate 
eigenfunctions $\psi_\I^\pm(Q)$ of the forms
\[
\psi_\I^+(Q) = \Theta(Q)Q^{-1/2+i\I/\hbar}
\]
and 
\[
\psi_\I^-(Q) = \psi_\I^+(-Q)
\]
respectively. We will concentrate on the solution $\psi_\I^+(Q)$,
which as we will now show represents bombardment of the equilibrium 
from the reactant side and is therefore the solution singled out in 
section~\ref{rcsol}.

The incoming flux is normalised as follows. Let the projection operator
$\hat{\Theta}$ have the $Q$-representation
\[
\hat{\Theta}\psi(Q) = \Theta(Q_0-Q)\psi(Q)
\]
so that it measures flux from  right to left in the $QP$-plane
across a section $\Sigma_R$ defined by $Q=Q_0$. Then
\[
\frac{1}{i\hbar}[\hat{\Theta},\Ih] 
= \frac{1}{2i\hbar}\left(\Qh[\hat{\Theta},\Ph]+[\hat{\Theta},\Ph]\Qh\right)
= -Q \delta(Q-Q_0)
\]
and sectional overlaps are of the form
\[
\brassket{\psi}{\Sigma_R}{\psi}
= -\frac{1}{i\hbar}\braopket{\psi}{[\hat{\Theta},\Ih]}{\psi}
= Q_0 |\psi(Q_0)|^2.
\]
The case $Q_0>0$ corresponds to incoming flux on the reactant side.
We then denote $\Sigma_R=\Sigma_R^\in$ and get
\[
\brassket{\psi_\I^+}{\Sigma_R^\in} {\psi_\I^+}
= Q_0 |Q_0^{-1/2+i\I/\hbar}|^2
=1.
\]
A section with $Q_0<0$ gives a flux in the incoming product channel 
and this vanishes for the state  $\psi_\I^+(Q)$, consistent with our 
interpretation of it as an incoming state in the reactant channel.

Outgoing fluxes are naturally measured in momentum representation
\[
\varphi(P) = \braket{P}{\psi}
\]
using projections of the form
\[
\hat{\Theta}\varphi(P) = \Theta(P-P_0)\varphi(P).
\] 
In this representation we have
\[
\frac{1}{i\hbar}[\hat{\Theta},\Ih] 
= \frac{1}{2i\hbar}\left([\hat{\Theta},\Qh]\Ph+\Ph[\hat{\Theta},\Qh]\right)
= -P \delta(P-P_0)
\]
and therefore
\[
\brassket{\psi}{\Sigma_P}{\psi} = P_0|\varphi(P_0)|^2.
\]
To complete the calculation we therefore need to evaluate
\begin{eqnarray*}
\varphi_\I^+(P) &=& \frac{1}{\sqrt{2\pi\hbar}}
\int_{-\infty}^\infty \e^{-iQP/\hbar}\psi_\I^+(Q)\d Q\\[6pt]
&=& \frac{1}{\sqrt{2\pi\hbar}}\left(\frac{|P|}{\hbar}\right)^{-1/2-i\I/\hbar}
\int_0^\infty \e^{-i\sigma q} q^{-1/2+i\I/\hbar}\d q \\[6pt]
&=& \frac{\hbar^{i\I/\hbar}}{\sqrt{2\pi}}
\e^{-i\sigma\pi/4+\sigma\pi\I/(2\hbar)}\Gamma\left(\ha+\frac{i\I}{\hbar}\right)
|P|^{-1/2-i\I/\hbar},
\end{eqnarray*}
where $\sigma$ is the sign of $P$. Note that 
$\varphi_\I^+(P)\sim{\rm const}\times|P|^{-1/2-i\I/\hbar}$ 
has the same dependence on its argument as found in the $Q$-representation,
which is to be expected since there is a symmetry between $\Qh$ and $\Ph$
in $\Ih$.

In either case, for flux calculations it suffices to know that
\begin{eqnarray*}
|\varphi_\I^+(P)|^2 
& = & \frac{\e^{\sigma\pi\I/\hbar}}{\e^{\pi\I/\hbar}+\e^{-\pi\I/\hbar}}
\frac{1}{|P|},
\end{eqnarray*}
where we have used
\[
\left|\Gamma\left(\ha+\frac{i\I}{\hbar}\right)\right|^2 
= \frac{\pi}{\cosh\pi\I/\hbar}.
\]
The $I$-flux across a section $\Sigma^\out$ defined by $P=P_0$ is therefore
\[
\brassket{\psi_\I^+}{\Sigma^\out}{\psi_\I^+} 
=\sigma \frac{\e^{\sigma\pi\I/\hbar}}{\e^{\pi\I/\hbar}+\e^{-\pi\I/\hbar}}
\]
where here $\sigma$ is the sign of $P_0$. This is positive in the 
reactants-out channel ($P>0$) and negative in the products-out channel
($P<0$) which,  when we remember that $\hat{\Theta}$ is defined
so that upward fluxes are positive, is consistent with  Figure~\ref{QPfig}.
We have therefore confirmed Equation (\ref{TR}).

\end{document}